# Digital Adaptive Carrier Phase Estimation in Multi-Level Phase Shift Keying Coherent Optical Communication Systems

Tianhua Xu, Tiegen Liu, Yimo Zhang
College of Precision Instrument and Opto-Electronics Engineering
Tianjin University
Tianjin, China
e-mail: xutianhua@tju.edu.cn; tianhua.xu@ucl.ac.uk

Gunnar Jacobsen, Jie Li
Networking and Transmission Laboratory
Acreo Swedish ICT AB
Stockholm, Sweden

Sergei Popov
School of Information and Communication Technology
Royal Institute of Technology
Stockholm, Sweden

*Abstract*—The analysis of adaptive carrier phase estimation is investigated in long-haul high speed n-level phase shift keying (n-PSK) optical fiber communication systems based on the one-tap normalized least-mean-square (LMS) algorithm. The close-form expressions for the estimated carrier phase and the bit-error-rate floor have been derived in the n-PSK coherent optical transmission systems. The results show that the one-tap normalized LMS algorithm performs pretty well in the carrier phase estimation, but will be less effective with the increment of modulation levels, in the compensation of both intrinsic laser phase noise and equalization enhanced phase noise.

## I. INTRODUCTION

More than 90% of the digital data is transmitted over optical fibers, to constitute the great part of the national and international communication infrastructures [1-6]. The performance of long-haul high speed optical fiber communication systems can be significantly degraded by the transmission impairments, such as chromatic dispersion (CD), polarization mode dispersion (PMD), laser phase noise (PN) and fiber nonlinearities (FNLs) [7-12]. With the full capture of the amplitude and the phase of optical signals, coherent optical detection and digital signal processing (DSP) allow the powerful equalization and mitigation of the communication system impairments in the electrical domain, and thus have become one of the most promising techniques for the next-generation optical fiber transmission networks [13-36]. Recently, some feed-forward and feed-back carrier phase estimation (CPE) algorithms have been proposed to compensate the phase noise from the laser sources [22-34]. Among these reported carrier phase estimation methods, the one-tap normalized least-mean-square (LMS) algorithm has been validated for compensating the laser phase noise effectively in the high speed coherent optical transmission systems [24-26].

Meanwhile, due to the interplay between the electronic dispersion equalization (EDC) and the laser phase noise, an effect of equalization enhanced phase noise (EEPN) has been generated and will seriously degrade the performance of long-haul optical fiber communication systems [37-42]. Some investigations regarding the equalization enhanced phase noise have been carried out in the single-channel, the wavelength division multiplexing (WDM), the orthogonal frequency division multiplexing (OFDM), the dispersion pre-distorted, and the multi-mode optical transmission systems [43-48]. The EEPN will distort the optical communication systems more severely with the increment of fiber dispersion, laser linewidths, modulation formats, and symbol rates [37,38,45]. Considering the impact of EEPN, the traditional analysis of the carrier phase estimation is not suitable any longer for the design of long-haul optical fiber transmission networks. Meanwhile, the requirement on laser linewidths will not be relaxed with the increment of signal symbol rates in communication systems. Therefore, it will be interesting to investigate the bit-error-rate (BER) performance in the one-tap normalized LMS carrier phase estimation algorithm, when the equalization enhanced phase noise is taken into account.

In our previous work, the analytical derivation for the one-tap normalized LMS carrier phase estimation method has been carried out based on the quadrature phase shift keying (QPSK) coherent optical transmission system [25,40]. It has been found that the one-tap normalized LMS equalizer behaves similar as the traditional differential detection in the carrier phase estimation for compensating both intrinsic laser phase noise and equalization enhanced phase noise in the QPSK coherent optical systems [25,40]. However, with the development of the optical fiber networks, and the increment of transmission data capacity, the QPSK modulation format cannot satisfy the demand of the high speed optical fiber communication systems any more. Therefore, the analysis on the carrier phase estimation approaches should also be updated correspondingly for the optical fiber transmission systems using higher-level modulation formats, such as *n*-level phase shift keying (*n*-PSK).

In this paper, a theoretical assessment for the carrier phase recovery using the one-tap normalized LMS algorithm in the *n*-PSK coherent optical transmission systems has been presented in detail. The analysis of the one-tap normalized LMS algorithm has been discussed, and the close-form expressions to predict the estimated carrier phase and the BER performance, such as the BER floor in the carrier phase recovery process, have also been described. It can be found that for the *n*-PSK optical transmission systems, the one-tap normalized LMS algorithm still shows a similar performance as the traditional differential carrier phase recovery. It can also be seen that the one-tap normalized LMS algorithm works very well in the carrier phase estimation in *n*-PSK optical fiber communication systems, but will be less effective with the increment of modulation levels, for compensating both intrinsic laser phase noise and equalization enhanced phase noise.

## II. LASER PHASE NOISE AND EQUALIZATION ENHANCED PHASE NOISE IN *n*-PSK TRANSMISSION SYSTEMS

In *n*-PSK coherent optical communication systems, the variance of the intrinsic phase noise from the transmitter (Tx) laser and the local oscillator (LO) laser can be expressed as follows [7,8],

$$\sigma^2_{Tx\_LO} = 2\pi(\Delta f_{Tx} + \Delta f_{LO}) \cdot T_S, \quad (1)$$

where $\Delta f_{Tx}$ and $\Delta f_{LO}$ are the 3-dB linewidths of the Tx laser and the LO laser respectively, and $T_S$ is the signal symbol period of the coherent transmission system. It can be found that the phase noise variance decreases with increment of the signal symbol rate $R_S = 1/T_S$.

However, due to the interaction between the dispersion and the LO laser phase noise, the noise variance of the equalization enhanced phase noise in the EDC based optical transmission systems can be expressed as follows, see in [37,40]

$$\sigma^2_{EEPN\_LO} = \frac{\pi \lambda^2}{2c} \cdot \frac{D \cdot L \cdot \Delta f_{LO}}{T_S}, \quad (2)$$

where $f_{LO}$ is the LO laser central frequency, which is usually equal to the Tx laser central frequency $f_{Tx}$, $D$ is the CD coefficient of the transmission fiber, $L$ is the transmission fiber length, and $\lambda = c/f_{Tx} = c/f_{LO}$ is the central wavelength of the optical carrier wave. It can be seen that the EEPN noise variance increases with the increment of symbol rate, which has an opposite feature compared to the laser phase noise.

## III. ONE-TAP NORMALIZED LMS CARRIER PHASE ESTIMATION IN *n*-PSK TRANSMISSION SYSTEMS

### A. Analysis of one-tap normalized LMS carrier phase estimation

As an adaptive feed-back algorithm, the transfer function of one-tap normalized LMS carrier phase estimation can be expressed as follows,

$$y(k) = w(k)x(k), \quad (3)$$

$$w(k+1) = w(k) + \frac{\mu}{|x(k)|^2} e(k) x^*(k), \quad (4)$$

$$e(k) = d(k) - y(k), \quad (5)$$

where $k$ is the symbol index, and $x(k)$ is the input symbol, $y(k)$ is the output symbol, and $w(k)$ is the tap weight of the one-tap normalized LMS equalizer, $e(k)$ is the carrier phase estimation error.

According to our previous work [40], the carrier phase estimation error can be expressed as the follows:

$$\Delta \phi = \phi_{k+1} - \phi_k. \quad (6)$$

Therefore, for the *n*-PSK coherent optical transmission systems, the demodulation part will not cause any errors, when we have

$$-\frac{\pi}{n} < \phi_{k+1} - \phi_k < \frac{\pi}{n}. \quad (7)$$

It is known that, for the Lorentzian distributed laser phase noise, $\phi_{k+1} - \phi_k$ follows a Gaussian distribution as the following expression:

$$f(x) = \frac{1}{\sqrt{2\pi}\sigma} \exp\left(-\frac{x^2}{2\sigma^2}\right), \quad (8)$$

where $\sigma^2$ is the variance of the laser phase noise, which can be calculated from (1).

Therefore, the symbol-error-rate (SER) in the *n*-PSK optical fiber transmission systems considering the laser phase noise can be calculated as follows:

$$P_{SER}(e) = \int_{-\infty}^{-\frac{\pi}{n}} f(x)dx + \int_{+\frac{\pi}{n}}^{+\infty} f(x)dx = erfc\left(\frac{\pi}{n\sqrt{2}\sigma}\right). \quad (9)$$

Therefore, the BER floor induced by the one-tap normalized LMS carrier phase recovery in the *n*-PSK optical fiber communication systems can be derived accordingly,

$$BER_{floor} = \frac{1}{\log_2 n} P_{SER}(e) = \frac{1}{\log_2 n} erfc\left(\frac{\pi}{n\sqrt{2}\sigma}\right) \quad (10)$$

where the noise variance $\sigma^2 = 2\pi(\Delta f_{Tx} + \Delta f_{LO}) \cdot T_S$, when only the intrinsic laser phase noise is considered.

It can be found that, for the *n*-PSK transmission systems, the close-form prediction for the BER floors in the one-tap normalized LMS carrier phase recovery algorithm also gives the same expression as the differential carrier phase recovery [7,40]. It means that the one-tap normalized least-mean-square carrier phase recovery in the *n*-PSK systems also behaves similar as the traditional differential carrier phase recovery.

### B. Influence of EEPN in the one-tap normalized LMS carrier phase estimation

When the EEPN is taken into account in the one-tap normalized LMS carrier phase estimation, we have the total noise variance in the optical fiber transmission system as the following expression,

$$\sigma_T^2 = \sigma_{Tx\_LO}^2 + \sigma_{EEPN}^2$$
$$= 2\pi T_S(\Delta f_{Tx} + \Delta f_{LO}) + \frac{\pi\lambda^2}{2c} \cdot \frac{D \cdot L \cdot \Delta f_{LO}}{T_S} \cdot \quad (11)$$

Therefore, considering the equalization enhanced phase noise, the BER floor in the one-tap LMS carrier phase recovery in *n*-PSK coherent optical communication systems can be evaluated as,

$$BER_{floor}^{PN+EEPN} = \frac{1}{\log_2 n} erfc\left(\frac{\pi}{n\sqrt{2}\sigma_T}\right). \quad (12)$$

## IV. RESULTS AND DISCUSSIONS

The results based on the above theoretical analyses for the one-tap normalized LMS carrier phase estimation is discussed in this section, by considering the influence from the laser phase noise and the EEPN.

The BER floors for different phase noise variances in the *n*-PSK coherent optical transmission systems are shown in Fig. 1, where the one-tap normalized LMS algorithm is employed for the carrier phase estimation.

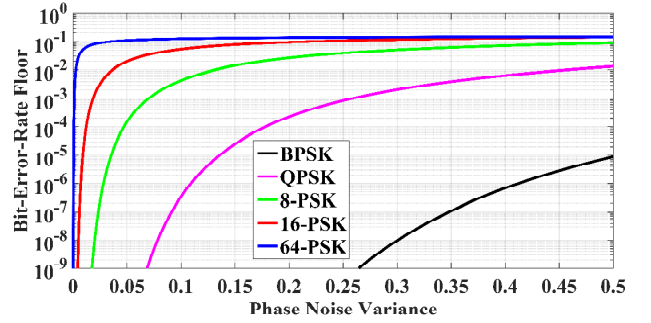

Fig. 1. BER floors versus different noise variances in the *n*-PSK coherent optical transmission systems using the one-tap normalized LMS carrier phase estimation.

It can be seen in Fig. 1 that the performance of coherent optical communication systems is degraded seriously by the phase noise (including the intrinsic phase noise and the equalization enhanced phase noise) with the increment of the phase noise variance. The effect is more significant for higher-level modulation formats.

As shown in Fig. 2, the BER floors for different laser linewidths in the *n*-PSK coherent optical fiber transmission systems have also been investigated using the one-tap normalized LMS algorithm for the carrier phase recovery. It can be found that the performance of optical communication systems is also degraded by the phase noise (in this case only intrinsic laser phase noise is considered) more seriously with the increment of the laser linewidths and the modulation levels.

Considering the influence of equalization enhanced phase noise, the BER floors for different transmission distances in the one-tap normalized LMS carrier phase estimation in the *n*-PSK coherent optical transmission systems are shown in Fig. 3, where both the Tx and the LO laser linewidths are set to 2 MHz. It can be seen that the performance of the optical fiber communication systems is degraded by the equalization enhanced phase noise significantly with the increment of the transmission distances, and the effect is more serious for higher-level modulation formats.

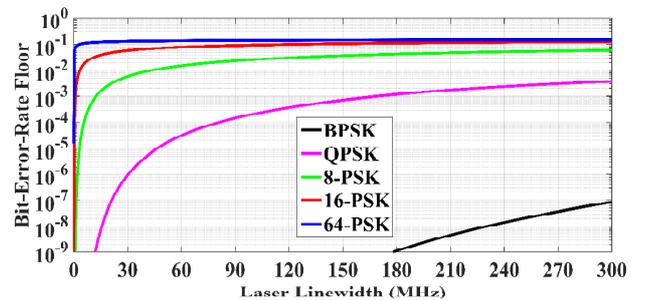

Fig. 2. BER floors versus laser linewidths in the *n*-PSK coherent optical transmission systems using the one-tap normalized LMS carrier phase estimation. The indicated linewidth value is the 3-dB linewidth for both the Tx and the LO lasers.

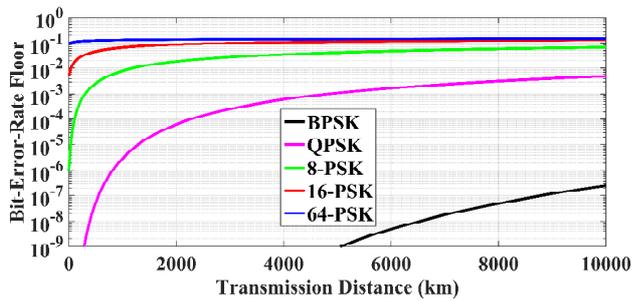

Fig. 3. BER floors versus transmission distances in the *n*-PSK coherent optical transmission systems using the one-tap normalized LMS carrier phase estimation, when equalization enhanced phase noise is considered. Both the Tx and the LO laser linewidths are 2 MHz.

## V. Conclusion

The theoretical evaluation of the carrier phase estimation using the one-tap normalized LMS algorithm in the *n*-PSK coherent optical transmission systems has been investigated. The close-form expressions for predicting the estimated carrier phase and the BER performance in the one-tap normalized LMS carrier phase estimation method have been presented in detail. For the *n*-PSK optical transmission systems, the one-tap normalized LMS algorithm still behaves similar as the traditional differential carrier phase recovery. It can be found that the one-tap normalized LMS algorithm works pretty well for the carrier phase estimation in the coherent *n*-PSK optical fiber communication systems, but will be less effective with the increment of modulation levels, in compensating both the intrinsic laser phase noise and the equalization enhanced phase noise.

## Acknowledgment

This work is supported in parts by EU project GRIFFON 324391, UK EPSRC project UNLOC EP/J017582/1, EU project ICONE 608099, and Swedish Vetenskapsradet 0379801.